# Pushing DSP-Free Coherent Interconnect to the Last Inch by Optically Analog Signal Processing

Mingming Zhang[†], Haoze Du[†], Xuefeng Wang, Junda Chen, Weihao Li, Zihe Hu, Yizhao Chen, Can Zhao, Hao Wu, Jiajun Zhou, Siyang Liu, Siqi Yan, and Ming Tang[*]

*Abstract*—To support the boosting interconnect capacity of the AI-related data centers, novel techniques enabled high-speed and low-cost optics are continuously emerging. When the baud rate approaches 200 GBaud per lane, the bottle-neck of traditional intensity modulation direct detection (IM-DD) architectures becomes increasingly evident. The simplified coherent solutions are widely discussed and considered as one of the most promising candidates. In this paper, a novel coherent architecture based on self-homodyne coherent detection and optically analog signal processing (OASP) is demonstrated. Proved by experiment, the first DSP-free baud-rate sampled 64-GBaud QPSK/16-QAM receptions are achieved, with BERs of 1e-6 and 2e-2, respectively. Even with 1-km fiber link propagation, the BER for QPSK reception remains at 3.6e-6. When an ultra-simple 1-sps SISO filter is utilized, the performance degradation of the proposed scheme is less than 1 dB compared to legacy DSP-based coherent reception. The proposed results pave the way for the ultra-high-speed coherent optical interconnections, offering high power and cost efficiency.

## I. Introduction

Rapidly advancing artificial intelligence (AI) will become a cornerstone of human society in the near future. The growing demand for AI training places greater pressure on the capacity provided by computing components. [1]. To further increase the component capacity, two practical approaches are available: scaling-up and scaling-out. Relying solely on scaling-up, constrained by Moore's Law, the scale growth of individual chips still cannot meet the demand. Therefore, scaling-out is necessary, which means involving the interconnection of separate chips and forming massive clusters. The scaling-out architecture is mainly enabled by high-speed, high-efficiency optical interconnect modules, which are extensively utilized in traditional data center networks (DCN) at present. For the past decade, hot-pluggable optical modules, e.g., small form-factor pluggable module (SFP), based on intensity modulation direct detection (IM-DD) have successful met the needs of the DCN market. This success can be attributed to the performance enhancements of optoelectronic devices and the relatively modest data traffic growth. However, in the scenario of AI training, the traditional optical interconnect modules fall short in many ways, such as speed, cost, density, power consumption, latency, and interconnect distance. More advanced and specialized solutions are required.

From an engineering perspective, the spatial density of power consumption becomes a significant limiting factor as switching capacity continues to increase. For a typical 128-port switch, where each port uses a standard 800G optical module with a power consumption of 15 W, the total power consumption on the front panel can reach 1.92 kW. Such high front-panel power consumption poses great challenges for both power supply and thermal management. Worse still, further increases in single-port speeds will inevitably lead to higher power consumption within an already compact and crowded space.

In the existing IMDD solutions, the three main approaches to increasing interface rates have each encountered bottleneck. *The first approach* is multi-lane parallel transmission. In next-generation 3.2T optical modules, 16 channels operating at 200 Gbps each will be required. Implementing parallel single mode (PSM)-based gray optics increases fiber costs and complicates the assurance of consistent insertion loss across different lanes. Alternatively, using coarse wavelength division multiplexing (CWDM)-based colored optics faces dispersion challenges. Even in the O-band, dispersion impairment at the edge channels is severe, significantly degrading performance over 2-km FR transmission. *The second approach* is to increase the baud rate, which is constrained by the ultra-wide bandwidth requirements of signal generators, drivers and modulators. Although time-domain or frequency-domain interleaving methods can artificially generate extra-high baud-rate signals, this approach offers less favorable power consumption improvements compared to multi-wavelength parallel transmission. *The third approach* is to employ high-order modulation formats, which requires power-hungry digital signal processing (DSP), particularly in retimed optics modules. As a result, IM-DD systems exhibit diminishing marginal returns as they approach next-generation high-speed interfaces.

By contrast, coherent systems offer several advantages, especially in terms of spectral efficiency. With this benefit, coherent systems can support equivalent capacity at only a quarter or even lower baud rate, reducing the dependence on high bandwidth devices and minimizing the need for numerous lanes. Despite these benefits, the high cost of narrow-linewidth lasers and the power consumption of DSP hinder the application of traditional coherent systems within data centers.

[M]ingming Zhang, Haoze Du, (†: equal contribution) Junda Chen, Weihao Li, Zihe Hu, Yizhao Chen, Can Zhao, Hao Wu, Siqi Yan, and Ming Tang (*: corresponding author) are with Huazhong University of Science and Technology, China; Xuefeng Wang is with Guangxi University, China; Jiajun Zhou and Siyang Liu is with JFS Laboratory, Wuhan, China.

Nevertheless, simplified coherent solutions have been developed to address these challenges. Recent research has shown, simplified coherent technology has significantly narrowed the power and cost gap when compared to IM-DD systems in short-reach applications. Among the simplified coherent receivers including complex-value direct detection (CV-DD) [2], [3], analog signal processing coherent detection (ASP-CD) [4], and self-homodyne coherent detection (SHCD) [5], the SHCD scheme has been verified by real-time experiment [6] and demonstrated as the most 'coherent-lite' scheme [7]. It is attributed to its less dependency on narrow-linewidth laser, lower complexity in DSP and full compatibility with the legacy coherent front-end. By matching the relative time delay (RTD) between the signal and the remotely delivered local oscillator (LO), the frequency-offset compensation (FOC) algorithm can be removed and the carrier phase recovery (CPR) can be significantly simplified. With the assistance of adaptive polarization controller (APC) and bidirectional transmission, the multiple-in-multiple-out (MIMO) algorithm for polarization demultiplexing can be also eliminated [8]. The highlight of those studies is the use of optically analog signal processing (OASP) to replace part of the traditional DSP, verifying the feasibility of substituting optical domain signal processing for digital domain processing. Compared to the high-bandwidth high-updating-speed DSP, those optical-domain processing methods show equal effectiveness with less bandwidth requirement and lower updating speed, leading to an increase in cost efficiency and a decrease in power and heat dissipation.

In this work, we experimentally study the DSP-free coherent communication systems with OASP. By introducing optical domain clock synchronization technology combined with injection locking, the digital clock recovery can be removed in baud-rate sampling configurations. Polarization demodulation can be solved in the optical domain by APC, and the self-homodyne coherent architecture can eliminate the frequency offset compensation and the carrier phase recovery algorithms. Leveraged by those techniques above, the sequential operating DSP flows have been recomposed into a parallel and decoupled form, and fully replaced by OASP. Finally, we experimentally demonstrate a DSP-free baud-rate sampling dual-polarization 64-GBaud QPSK signal reception with no error (pre-FEC BER < 1e-6) in back-to-back (BtB) scenario and with a pre-FEC BER of 3.6e-6 in a 1-km fiber link.

## II. Optically Analog Signal Processing

When propagating along the short-reach fiber links, the full-field coherent detection can be vulnerable to many stochastic perturbations and determined impairments. The **stochastic perturbations** mainly include:
1) sampling clock jitter between the transmitter side and receiver side.
2) stochastic rotation and dispersion of polarization caused by the random birefringence of the optical fiber.
3) phase jitter and frequency offset between the laser source and LO.

Except those perturbations, the signal is also influenced by **determined impairments** such as, frequency-domain distortions, and optical-domain chromatic dispersion. In traditional coherent detection systems, especially in the long-haul scenarios, perturbations and impairments are highly coupled and mixed. To recover and demodulate the original signal, DSP is usually necessary. The DSP must operate under a symbol/block-wise basis to ensure the robustness and performance of communication system. Nevertheless, that is cumbersome in short-reach scenarios, where the perturbations and impairments are lite enough to be compensated in optical domain.

The DSP-based and OASP-based coherent receiver schemes are illustrated and compared in Fig.1. In traditional DSP flows, the resampling and clock recovering (RS & CR) algorithms (a) are used to compensate the clock jitter. The MIMO adaptive equalization (AEQ) algorithms (b) are used to eliminate the inter-polarization interferences and demultiplex the dual-polarization signal. The frequency-offset compensation and carrier phase recovery (FOC & CPR) algorithms (c) are used to eliminate the influence of the phase noise of lasers. Those algorithms operate sequentially in flows, which means the former sections' results are critical to the following ones. However, those compensation operations are not necessarily in queue within OASP-based architecture. In optical domain, they can be cooperating parallelly: those three function blocks can be respectively substituted by optical clock synchronization unit (a), adaptive polarization controlling unit (b), and remotely delivered self-homodyne LO recovery unit (c).

### A. Optical clock Synchronization

In ultra-high-speed serial communications, the electrical clock delivering lanes are considered susceptible to interference and prone to significant attenuation. Consequently, the clock synchronization is usually dependent on clock and data recovery (CDR) module, which acquires clock information by signal waveforms themselves. However, in coherent systems, the phase noise and high-order modulation formats increase the difficulty and complexity of CDR. Meanwhile, the digital CDR hinders the realization of DSP-free baud-rate sampling coherent receivers.

The recent progress of optical clock distribution shows incredible precision and robustness [9]. Luckily, this optical clock distribution architecture is endogenously embedded into SHCD systems. The picosecond-RTD aligned twin-links ensures the simultaneous arrival of reference clock information and the data signal waveform.

In the proposed architecture, the optical reference clock signal is loaded into the LO light on the transmitter side using a simple sinusoidal intensity modulation. And the reference clock waveform is detected by a photodiode (PD) on the receiver side. The frequency of the reference clock should be several gigahertz, to reduce the influence on coherent detection after LO recovery.

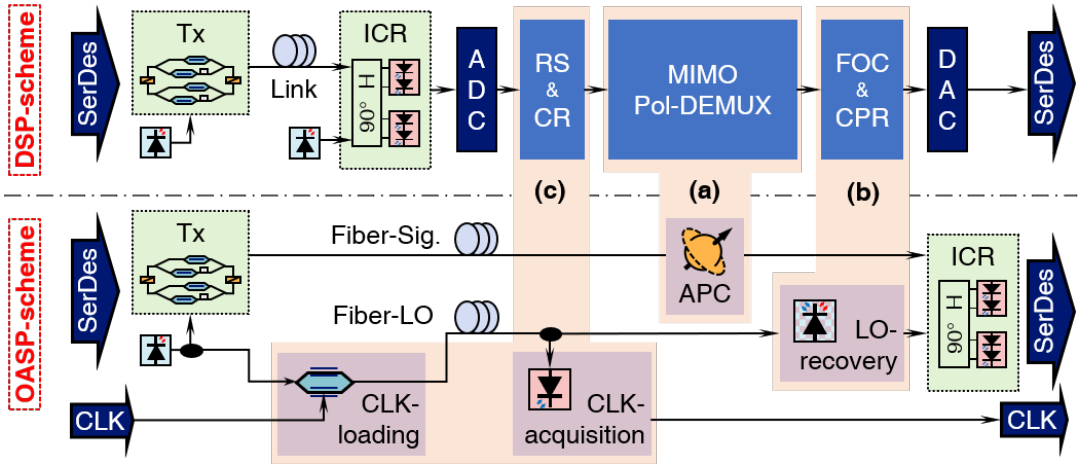

Fig. 1. The proposed OASP-based self-homodyne coherent receiver and the comparison against classical DSP scheme.

*B. Adaptive polarization controlling and demultiplexing*

In short-reach scenarios, polarization perturbation primarily involves polarization rotation, with differential delay between polarizations being negligible. As a result, the demultiplexing can be implemented by a unitary polarization controller, rather than AEQ FIR filters. High-speed electrically controlled silicon-photonic APC is fully able to track the polarization variation through the link [10]. Our recent work has increased the tracking speed to 100 krad/s [11]. By introducing the feedback circuit and controlling algorithms, fully optical domain polarization demultiplexing can be realized [12]. Remarkably, with optical domain demultiplexing, the feedback circuit bandwidth is significantly narrower than that of the main signal processing circuit. Also, the full silicon-based APC devices are highly compatible with integration to the silicon-based ICRs.

*C. Remotely delivered self-homodyne LO recovery*

Different from classical coherent architectures, the SHCD receivers utilize the LO light delivered from the transmitter side. Due to the laser homology of LO and signal, there is no frequency offset and tiny phase noise. It has been proved that, when the RTD is well matched, the phase noise of SHCD can be significantly suppressed even with a non-narrow-linewidth laser [13]. Due to the extremely low phase noise, the absolute phase of the signal can remain stable for a long period, making the CPR algorithm no longer necessary. The only drawback is that the link attenuation of the LO power decreases the sensitivity of receivers. On the other hand, as discussed in Section II.A, regarding the optical clock, the optical clock has been embedded into the LO, leading to significant LO power fluctuations.

To address both issues, an injection-locked DFB laser (IL-DFB), which is a commercial large-linewidth DFB laser without output isolator, is used as a high-efficiency LO recovery module [14]. It can remove the clock signal interference from the LO while simultaneously amplifying the LO power and preserving the low phase noise characteristics [15], due to its selective frequency amplification property. To avoid the injection intensity induced phase distortion of the IL-DFB, the optical clock frequency is determined to be at least around 2.5 GHz, by experiments.

Above all, by employing all-optical tracking and OASP techniques, almost all the stochastic perturbations can be well mitigated. Further, each type of perturbation can be treated as uncoupled and dependent in short-reach scenarios, thus making the OASP-based compensation more effective. Finally, the baud-rate sampled DSP-free SHCD communication system can be realized.

### III. EXPERIMENT VERIFICATION OF OASP AND DISCUSSIONS

The experiment set-up is shown in Fig. 2. The system includes 3 parts, signal path, reference clock path and LO path carrying clock signal. APC is applied in signal path to optically demultiplex the dual-polarization signal. Clock signal is loaded on the LO by a Mach-Zehnder modulator (MZM) in the transmitter end and then remotely delivered to the receiver end. After that, an IL-DFB is used to erase the clock signal and regenerate the LO.

The key point of this system is the refence clock path which is denoted as blue line in the figure. The reference clock signals are derived from the same arbitrary waveform generator (Keysight, M8196A), which acts as a clock source. Based on the homologous reference clock signals, the transmitter and receiver can be precisely synchronized. The frequencies of each reference clocks are dependent on many factors, including devices restrictions, timing jitter and LO regeneration. The clock synchronization of transmitting signal generator is achieved by phase-locked loop in the AWG (Keysight, M8194A), which supports external clock frequencies in the range of 200 ~ 300 MHz. Meanwhile, the oscilloscope at the receiver can only support 10-MHz input clock signal. On the other hand, the LO sidebands introduced by clock signal will be removed by the IL-DFB. When the sidebands are too close to the LO, the performance of LO recovery will be degraded. To mitigate additional time jitter, the frequencies of the reference clocks must be selected as integer multiples of a fundamental frequency. Therefore, the reference clock sent to the transmitter is set to 200 MHz, and the reference clock loaded on the optical LO is set to 4 GHz, which will be down-converted to a 10-MHz electrical reference clock by a 400-fold analog divider. Even after propagating through the fiber link, the 4-GHz optical clock signal exhibits a root-mean-squared (RMS) time jitter of 3.9 ps, shown in insert figure named as "Electrical Waveform (O-CLK)". Variable optical delay line (VODL) is applied before PD to manually adjust the phase of optical clock, to fine-tune the oscilloscope's sampling time point.

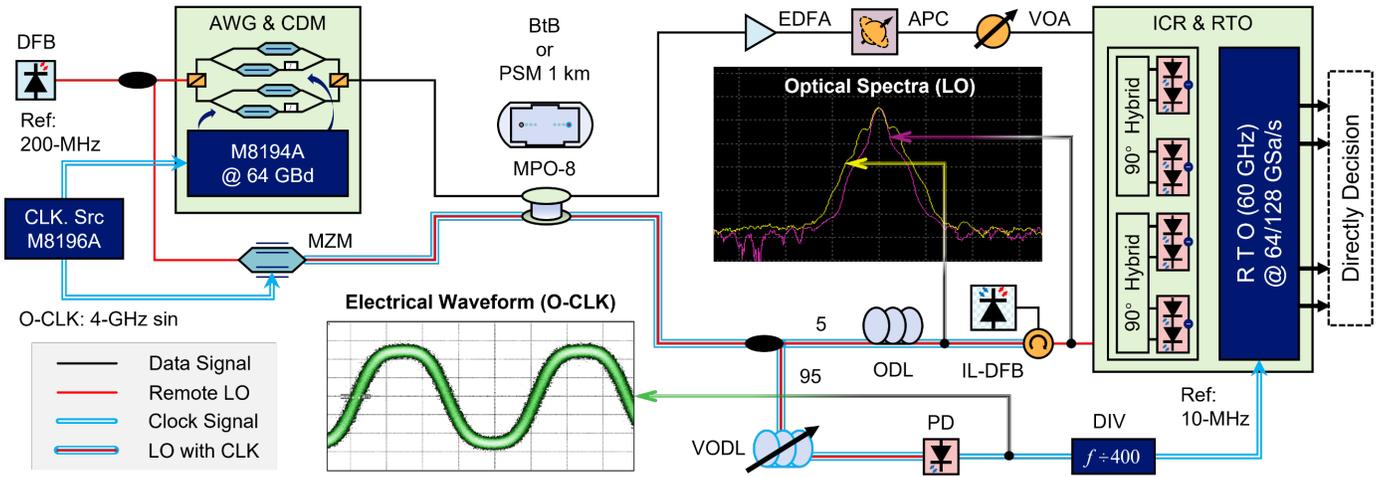

Fig. 2. Experiment setup of OASP-based SHCD system.

For optical path in the SHCD system, signal and LO come from the same DFB laser and are separated by a 50:50 coupler. LO path is indicated by red line. After passing through the MZM, LO carries the optical clock signal into the optical fiber link. Before inputting the SHCD receiver, the LO is divided by a 95:5 coupler. The majority of LO is collected by a PD and the rest part of LO is injected into the IL-DFB laser. The use of the majority of the power for clock recovery is for the low sensitivity of the PD in contrast to the high sensitivity of IL-DFB. The input power of IL-DFB is about -18 dBm, while the output power is 13 dBm. Illustrated in insert figure named as "Optical Spectra (LO)", the free-running IL-DFB will be forced to synchronize with the injected remote LO light, lasing at the same frequency and phase, while suppressing the sidebands induced by the optical clock. Additionally, the RTD between the signal and the LO is well matched by the optical delay line (ODL), yielding a residue of about 10 ps. Signal path is indicated by black line. Dual polarization 64-GBaud QPSK/16-QAM signals are generated by the arbitrary waveform generator (AWG, Keysight M8194A), modulated by the coherent driver modulator (CDM, ID-Photonics), and launched into the fiber link. To avoid the RTD among different lanes, a 1-km parallel single mode fiber link with an 8-channel multi-fiber push-on (MPO-8) connector is employed. After propagated through the fiber link, the dual-polarization signal is demultiplexed by an APC. An erbium-doped fiber amplifier (EDFA) is used to compensate the 10-dB insertion loss of the APC. A variable optical attenuator (VOA) is employed to facilitate a broader sweep of the received optical power (ROP).

The signals and the RTD-matched recovered LO are coherently detected in the integrated coherent receiver (ICR). Finally, the data streams are captured by the real-time oscilloscope (RTO, Keysight UXR0594AP). Leveraging the precisely clock synchronizing, polarization demultiplexing and self-homodyne detecting, the 64-GBaud data streams can be directly received and decided by baud-rate sampling ADCs without any post-DSP. As a result, the BER can be measured in a pseudo real-time manner, even by an offline setup. Due to the feature of real-time BER measurement, the impact of time jitter between the transmitter and receiver can be analyzed. Sweeping the VODL in the optical clock path, the sampling clock phase can be fine-tuned. Illustrated in the top of Fig.3, the impact of clock phase offset results in the degradation of real-time BER. When applying 64-GBaud QPSK signal, the time jitter within 4 ps is tolerable under the 1E-4 FEC threshold. This also indicated that the synchronized clock is sufficiently stable for the system performance test.

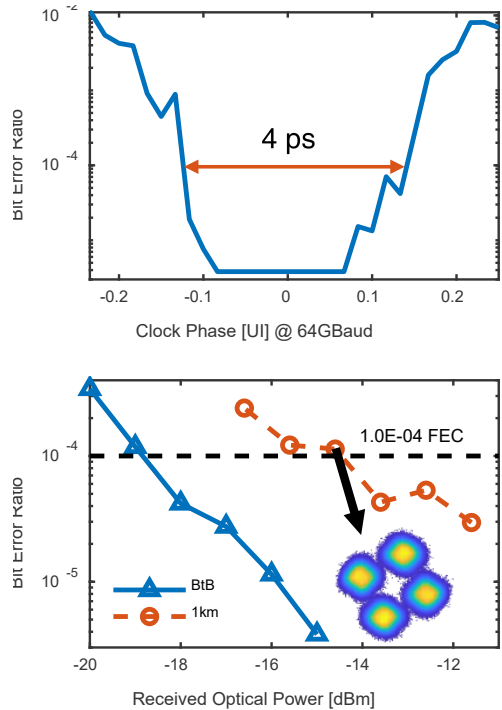

Fig. 3. (Top) The impact of the sampling clock phase offset on the system performance. (Bottom) The performance of the back-to-back and the 1-km scenarios.

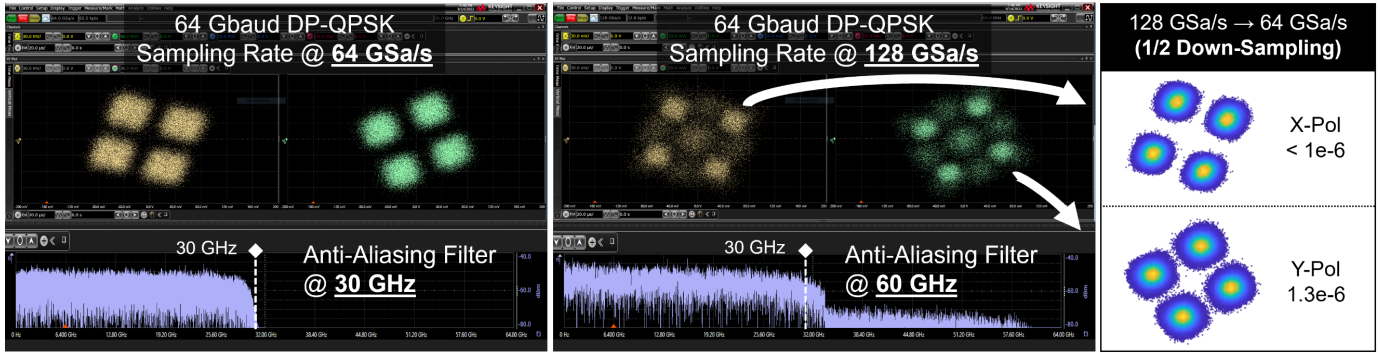

**Fig. 4.** The real-time constellations captured in the oscilloscope's screen. (Left) The baud-rate sampling results. (Right) The twice-baud-rate sampling results and the 1/2 down-sampled results.

The performances of the BtB and 1-km fiber link scenarios are illustrated at bottom of Fig. 3. To meet the pre-FEC BER threshold of 1E-4, the minimum ROP is respectively -19 dBm and -15 dBm, at BtB and 1-km link. There is a penalty of approximately 4 dB between the two scenarios, mainly because of the fiber dispersion at the wavelength of 1550 nm. This impairment can be mitigated by moving the wavelength to O-band at 1310 nm.

To demonstrate the pseudo real-time constellations, the sampling rate of RTO should be set at the same value of the signal's baud rate, at 64 GSa/s. After automatically demultiplexing the polarization and finely tuning the sampling clock phase, the constellation of the received coherent signal can be directly displayed as X-Y plot mode on the RTO's screen, without any DSP. The baud-rate sampling results are shown in the left subplot of Fig. 4. Due to the high sensitivity of the coherent structure's phase to environmental variations, the constellations exhibit very slow rotation, compared with that induced by laser's phase noise. The stable yet slowly rotating constellations of the 64-GBaud QPSK signal are clearly observable. However, when operating at a sampling rate of 64 GSa/s, the waveform capture encounters inter-symbol interference (ISI) due to the enforced 30-GHz anti-aliasing filter. To mitigate this, setting the sampling rate to twice the baud rate, or 128 GSa/s, with a subsequent 1/2 down-sampling step effectively enables single-baud-rate sampling. This method differs from conventional resampling based on FIR filters, as it provides equivalent results with potentially reduced computational overhead. The real-time constellations and down-sampled results are respectively shown in the right subplots of Fig. 4. Each frame lasts 15.6 μs.

Though the fully OASP scheme operates effectively in pseudo real-time tests, the performance of the system can be increased by highly simplified DSPs. To further discuss the performance differences, three signal processing methods with different complexity are demonstrated respectively in Fig. 5. *The first column* shows real-time constellations captured by the oscilloscope. The pre-FEC BER of the QPSK signal can reach below 1E-6, approaching an error-free level. Additionally, the pre-FEC BER of the 16-QAM signal can reach 2.2E-2, remaining within the SD-FEC threshold. *The second column* shows the constellations and SNRs under OASP with no DSP except for down-sampling. Furthermore, an ultra-simplified

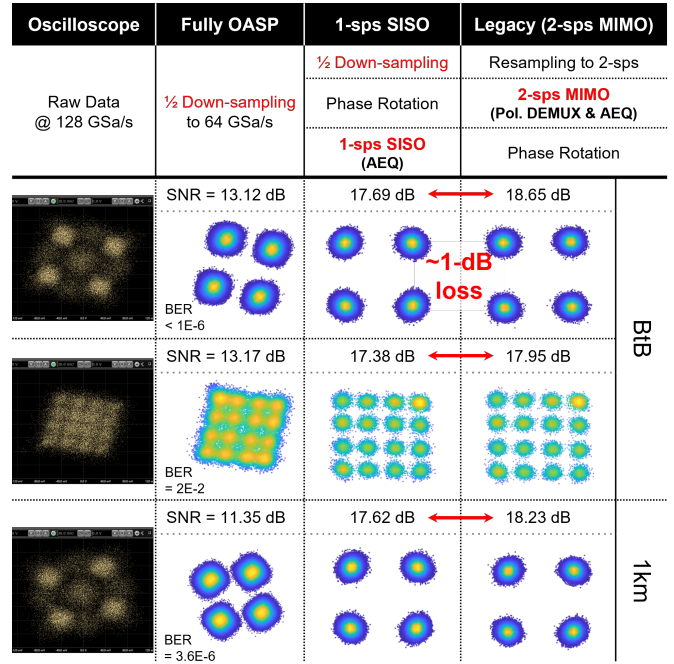

**Fig. 5.** The raw real-time constellations and comparison among three different signal processing schemes.

DSP is adopted to compensate the residual frequency-domain impairments, which is shown in *the third column*. Following the down-sampling and a simple phase rotating, the 1-sample per symbol single-in single-out (1-sps SISO) compensates most of the residue impairments. Compared with the legacy DSP (based on 2-sps MIMO), which is shown in *the last column*, the SNR losses between them are less than 1-dB. Even propagated through 1-km fiber link, the performance degradation is negligible compared with the BtB case. It is important to note that the 1-sps SISO equalization mentioned above is possible to be facilitated by the subsequent powerful electrical SerDes components, which is the case in the recently discussed LPO initiative.

## IV. CONCLUSION

In summary, this article proposes a DSP-free solution for achieving high-speed coherent short-reach interconnections in data centers. By employing all optical analog signal manipulations — including optical clock synchronizing, adaptive polarization controlling, and self-homodyne coherent

detection — the first 64-Gbaud DSP-free QPSK/16-QAM reception has been experimentally realized using baud-rate sampling ADCs. This work successfully demonstrates that the algorithmic complexity of the coherent architectures can be reduced to a level comparable to that of IM-DD.

ACKNOWLEDGMENT

This work was supported by the National Natural Science Foundation of China (#62225110), the Major Program (JD) of Hubei Province (#2023BAA001-1).

BIOGRAPHIES

**Mingming Zhang** (carlzhang.hust@gmail.com) is currently a Ph.D. candidate at Huazhong University of Science and Technology, China.

**Haoze Du** (duhaoze@hust.edu.cn) is currently a Ph.D. candidate at Huazhong University of Science and Technology, China.

**Xuefeng Wang** (1124743493@qq.com) is currently a researcher at School of Computer and Electronic Information, Guangxi University, China.

**Junda Chen** (chenjd@hust.edu.cn) is currently a postdoctoral researcher at School of Optical and Electronic Information, Huazhong University of Science and Technology, China.

**Weihao Li** (weihaoleecn@qq.com) received his Ph.D. at Huazhong University of Science and Technology, China.

**Zihe Hu** (454890179@qq.com) is currently a Ph.D. candidate at Huazhong University of Science and Technology, China.

**Yizhao Chen** (2213577836@qq.com) received his Ph.D. at Huazhong University of Science and Technology, China.

**Can Zhao** (zhao_can@hust.edu.cn) is currently a researcher at Huazhong University of Science and Technology, China.

**Hao Wu** (wuhaoboom@hust.edu.cn) is currently a researcher at Huazhong University of Science and Technology, China.

**Jiajun Zhou** (zhoujiajun@jfslab.com.cn) is currently working at JFS Laboratory, Wuhan, China.

**Siyang Liu** (liusiyang@jfslab.com.cn) is currently working at JFS Laboratory, Wuhan, China.

**Siqi Yan** (siqya@hust.edu.cn) is currently an associate professor at School of Optical and Electronic Information, Huazhong University of Science and Technology, China.

**Ming Tang** (tangming@mail.hust.edu.cn) is currently a professor at School of Optical and Electronic Information, Huazhong University of Science and Technology, China.